\documentclass[
reprint,
superscriptaddress,
amsmath,
amssymb,
aps,
pra,
]{revtex4-2}

\usepackage{float}
\usepackage{graphicx}% Include figure files
\usepackage{dcolumn}% Align table columns on decimal point
\usepackage{bm}% bold math
\usepackage{xcolor}
\usepackage{gensymb}
\begin{document}

%\preprint{APS/123-QED}

% \title{Exploring core-shell design for levitated nanocryostats}
% \title{Exploring inert shell coating for lanthanide based nanocryostats with application in levitodynamics}
% \title{Exploring inert-shell coating for enhanced laser refrigeration: application in levitated optomechanics}
% \title{Exploring inert-shell coating for enhanced laser refrigeration of nanoparticles:\\ application in levitated optomechanics}
\title{Inert shell coating for enhanced laser refrigeration of nanoparticles:\\ application in levitated optomechanics}

\author{Cyril Laplane}
    \email{cyril.laplane@mq.edu.au}
    \affiliation{Sydney Quantum Academy, Sydney, NSW 2006, Australia}
    \affiliation{School of Mathematical and Physical Sciences, Macquarie University, NSW 2109, Australia}
    \affiliation{ARC Centre of Excellence for Engineered Quantum Systems, Macquarie University, NSW 2109, Australia}
\author{Peng Ren}
    \affiliation{School of Engineering, Macquarie University, NSW 2109, Australia}
    \affiliation{ARC Centre for Nanoscale BioPhotonics, Macquarie University, NSW 2109, Australia}
\author{Reece P. Roberts}
    \affiliation{School of Mathematical and Physical Sciences, Macquarie University, NSW 2109, Australia}
    \affiliation{ARC Centre of Excellence for Engineered Quantum Systems, Macquarie University, NSW 2109, Australia}
\author{Yiqing Lu}
    \affiliation{School of Engineering, Macquarie University, NSW 2109, Australia}
    \affiliation{ARC Centre for Nanoscale BioPhotonics, Macquarie University, NSW 2109, Australia}
\author{Thomas Volz}
    \affiliation{School of Mathematical and Physical Sciences, Macquarie University, NSW 2109, Australia}
    \affiliation{ARC Centre of Excellence for Engineered Quantum Systems, Macquarie University, NSW 2109, Australia}

\date{\today}% It is always \today, today,

\begin{abstract}
We report on a study exploring the design of nanoparticles that can enhance their laser refrigeration efficiency for applications in levitated optomechanics. In particular, we developed lanthanide-doped nanocrystals with an inert shell coating and compared their performance with bare nanocrystals. While optically levitated, we studied the refrigeration of both types of nanoparticles while varying the pressure. We found that the core-shell design shows an improvement in the minimum final temperature: a fourth of the core-shell nanoparticles showed a significant cooling compared to almost none of the bare nanoparticles. Furthermore, we measured a core-shell nanoparticle cooling down to a temperature of 147 K at 26 mbar in the underdamped regime. Our study is a first step towards engineering nanoparticles that are suitable for achieving absolute (centre-of-mass and internal temperature) cooling in levitation, opening new avenues for force sensing and the realization of macroscopic quantum superpositions.
\end{abstract}

\maketitle

%\tableofcontents

\section{\label{sec:level1}Introduction}
The field of levitated optomechanics - levitodynamics - presents a new paradigm for optomechanics with the promise of ultrahigh Q without the need for a resonator. The levitation of mesoscopic particles has recently established itself as an exquisite platform for force sensing with the potential of quantum advantage \cite{gonzalez-ballestero_levitodynamics_2021}. These archetypical isolated mechanical oscillators have very low coupling to the environment, with their linewidth essentially limited by the quality of the vacuum \cite{arita_coherent_2020,pontin_ultra-narrow_2019} and 
they have recently entered the quantum realm \cite{piotrowski_simultaneous_2023,tebbenjohanns_quantum_2021,magrini_real-time_2021,ranfagni_vectorial_2021}. The current state-of-the-art still presents limited coherence times which remains one of the limiting factors for the ultimate goal of matter-wave interferometry with these massive levitated particles. Ultimately one of the mechanisms responsible for decoherence is the internal temperature of the levitated particle which can reach several thousands of Kelvin even for small motional temperatures \cite{bateman_near-field_2014}. In this regime, one source of decoherence comes from blackbody radiation \cite{chang_cavity_2010,Wan2016}, limiting experiments in the quantum regime without any internal cooling mechanisms \cite{bose_spin_2017,frangeskou_pure_2018}. It is also interesting to note that even for a moderate vacuum, in the regime where damping mainly comes from collisions with the surrounding gas, having a colder levitated oscillator can in principle increase its Q factor since gas viscosity will increase with temperature \cite{millen_nanoscale_2014,rahman_laser_2017}. A higher Q at a moderate vacuum should in principle help to cool down the centre-of-mass (COM) motion.\\
The internal temperature of an optically  levitated object depends on its absorption cross-section, the laser wavelength and irradiance. This is one of the reasons why to date most successful experiments have used SiO$_2$ nanoparticles as this material exhibits low absorption cross-section at typical trapping wavelength (1064nm and 1550nm). Furthermore, the synthesis of SiO$_2$ nanoparticles is a mature technology with high yield and uniformity, which make them a prime choice to achieve repeatability in experiments. Levitodynamics experiments with optically active (i.e. absorbing) nanoparticles have been limited to moderate vacuum pressure P $>$ 1 mbar \cite{conangla_extending_2020}. Controlling the internal temperature of a levitated object in vacuum is a challenging task and requires contactless cooling using electromagnetic radiation.\\
Rare-earth ion doped crystals are one of the few materials enabling laser refrigeration of solids through anti-Stokes fluorescence \cite{Melgaard2016,rahman_laser_2017}, where the energy of the emitted radiation is greater than the energy of the absorbed light. A record low temperature of 91 K \cite{Melgaard2016} obtained with a 10$\%$ Yb$^{3+}$:LiYF$_4$ crystal  was only possible thanks to careful material engineering, i.e. finding the best crystalline host as well as growing an ultrahigh-quality crystal.
The laser refrigeration of optically trapped mesoscopic crystals in liquids (water and D$_2$O) have been demonstrated using 10$\%$ Yb$^{3+}$:LiYF$_4$ \cite{roder_laser_2015} and $\beta$-10$\%$ Yb$^{3+}$:NaYF$_4$ nanowires \cite{zhou_laser_2016,ortiz-rivero_laser_2021} with refrigeration of $\Delta$T $\approx$ -15K, -9K and -6K. In 2017, A. T. M. Anishur Rahman and P. F. Barker reported laser refrigeration in optical levitation with a record low temperature of 130 K and an average internal temperature of 167 K \cite{rahman_laser_2017}. It is important to note that in this work, a top-down approach was used to produce the nanocrystals (NC). 
\begin{figure*}
	% \centering
    \includegraphics[width=1\textwidth]{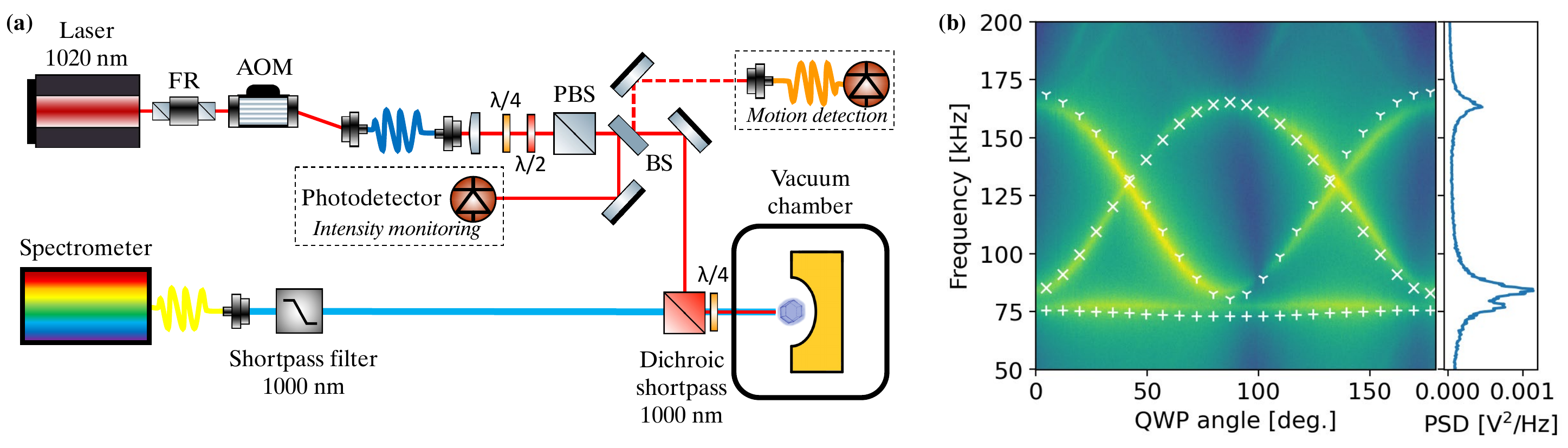}
    \caption{Experimental setup. (b) Oscillator spectroscopy of a levitated 10$\%$Yb$^{3+}$:NaYF$_4$ nanocrystal at P=7 mbar. The white markers show the simulated trap frequencies for the different translational degrees of freedom of the levitated particle.}
	\label{fig:exp_setup_fig}
\end{figure*}
The sample was obtained by milling down a bulk crystal which resulted in a wide variety of shapes and cooling efficiency. Finally in 2021, $\alpha$-10$\%$ Yb$^{3+}$:NaYF$_4$ synthesized levitated nanocrystals, grown with an hydrothermal process, have been cool down to an average temperature of 252 K (lowest 241 K) \cite{luntz-martin_laser_2021}.
It remains paramount to develop synthesis techniques that will yield high uniformity and quality of nanoparticles. Lanthanides doped in Sodium Yttrium fluoride matrices have been extensively studied for upconversion imaging \cite{rosal_upconversion_2019}. For this application, nanomaterials requirements such as high brightness and low phonon energies are similar to those for anti-Stokes cooling. It thus presents an interesting avenue to explore the same nanoengineering techniques to improve the performance and quality of our nanocryostats.
We here investigate the potential of Yb$^{3+}$:NaYF$_4$ nanocrystals with an inert outer shell. This shell reduces non-radiative losses at the surface which will help to maximise the quantum yield and hence the cooling efficiency. We performed fluorescence and oscillator spectroscopy of particles with and without an inert shell and we used a simple thermodynamic model to simulate the internal temperature of the levitated nanoparticles. Thanks to fluorescence thermometry at different pressure we can then assess the cooling properties of the levitated particle. Although of particular interest to the field of levitodynamics, this work also shows the potential of a bottom-up nanoengineering approach for optimising optical cryocooling materials. Ultimately it allows for more versatility in the study of material, crystal phase, morphology, design and dopant concentration compared to a bulk crystal approach that has stringent requirements on the purity/quality of the crystal.

\section{Methods}
\subsection{Experimental setup}
We optically trap $\beta$-phase 10$\%$Yb$^{3+}$:NaYF$_4$ and 10$\%$Yb$^{3+}$:NaYF$_4$@NaYF$_4$ (with an inert shell) nanocrystals in our experimental setup shown in Fig.\ref{fig:exp_setup_fig}a. Our trapping and cooling laser is tuned to 1020 nm where the cooling efficiency should be maximized \cite{Melgaard2016}. To create the optical tweezer, we have designed a gold-coated parabolic mirror (as described in \cite{vovrosh_parametric_2017}) with an effective numerical aperture (NA) close to 0.99, as confirmed by the oscillator spectroscopy (see Fig. \ref{fig:exp_setup_fig}b). The parabolic mirror trap presents several advantages particularly suited for this study: first the reflection coefficient of gold is relatively constant for the range of wavelength of interest, second the mirror presents virtually no chromatic aberrations compared to microscope objectives. This is particularly suited for multi-wavelength addressing of the levitated nanoparticle. The laser power was set to $\sim$ 77 mW which gives an excitation irradiance of 22.8 MW/cm$^{2}$ well above saturation.\\
The fluorescence from the nanocrystal is filtered through a 1000nm dichroic mirror and a 1000nm shortpass filter before coupled into a single-mode fibre connected to the spectrometer. The light scattered by the levitated nanocrystal is collimated by the parabolic mirror and collected through a multimode fibre which is sent to a photodiode. By adjusting the coupling in the fibre we can collect part of the unscattered laser light, hence we monitor the motion of the nanoparticle through a homodyne detection scheme \cite{rashid_wigner_2017}\\
The nanoparticles are first diluted in ethanol and sonicated for 90 min before being loaded into the trap at ambient pressure using an ultrasonic nebulizer. We then monitor the temperature of the nanoparticles while decreasing the pressure in the vacuum chamber.\\
At pressure P $\lesssim$ 20 mbar, the levitated oscillator enters the underdamped regime and we can perform spectroscopy (see Fig.\ref{fig:exp_setup_fig}b) by varying the ellipticity of the polarization of the trapping laser using a quarter-wave plate. We can approximate the NC as discs of the same diameter and thickness and by adapting an optical tweezer computational toolbox \cite{Lenton2020}, we can simulate the dynamics of the NC in our trap while varying the polarization (see white crosses in Fig. \ref{fig:exp_setup_fig}b). This allows us to confirm the properties of the parabolic mirror such as its NA as well as understanding the shape, size \cite{rademacher_measurement_2022} and orientation of the NCs in the trap.
\begin{figure*}
	\centering
    \includegraphics[width=1\textwidth]{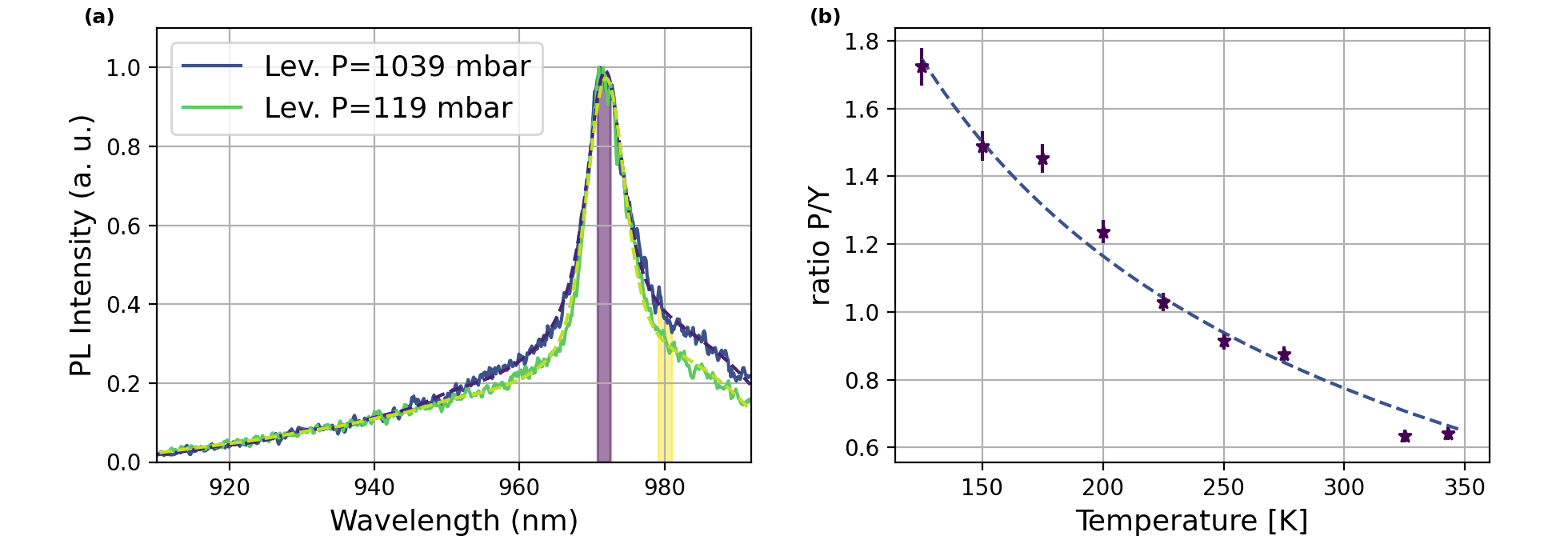}
    \caption{Photoluminescence spectra of a core-shell levitated particle at two different pressure. We estimate the temperature at 119 mbar to be 210 $\pm$ 1 K. The purple and yellow regions are the two transitions used for ratiometric thermometry. On the right side, we plot the ratio P(urple)/Y(ellow) measured in our cryostat at known temperatures to check the choice of the two transitions as a good thermometer. The blue dashed curve is a fit according to Eq.\ref{model_dI}.}
	\label{fig:spectrum_calib_cryo}
\end{figure*}
We take care that for the fluorescence spectroscopy, the polarization is always oriented to a particular optical axis of the nanocrystal thus ensuring a fixed absorption coefficient \cite{rahman_laser_2017}.
In the current article, we limit our study to trapping (and cooling) using linearly polarized light. This means that we prevent the libration and/or rotation of the nanocrystal \cite{rashid_precession_2018} in the trap which could affect its gas thermalization rate \cite{millen_single_2018}.

\subsection{Thermometry}
From the photoluminescence (PL) spectra we choose to evaluate the temperature of the NC through ratiometric thermometry \cite{patterson_measurement_2010,roder_laser_2015,rahman_laser_2017}. The ratio of PL intensities at two different wavelengths depends on the difference in populations between the energy levels involved. We can thus infer the temperature if we assume that the spectral irradiance for particular emission bands (purple and yellow in Fig.\ref{fig:spectrum_calib_cryo}) is given by a Boltzmann distribution. Looking at the intensity ratio of two different transitions (ab and cd) in the photoluminescence spectrum and assuming thermal equilibrium we have:
\begin{equation}
    R = \frac{I_{ab}}{I_{cd}} = Ae^{-\frac{E_{ab}-E_{cd}}{k_BT}}
    \label{model_dI}
\end{equation}
By measuring this ratio at a known temperature, we can calibrate the temperature measurement. In our experiment, we assume that the particle is at equilibrium with the gas temperature at ambient pressure, a condition which is confirmed by our simulations (see Results). Hence if $R_{1bar}$ is the measured ratio at $T_{1bar}$, the temperature $T$ at a pressure $P<1$ bar is given by:
\begin{equation}
    \frac{1}{T} = \frac{1}{T_{1bar}} + ln(\frac{R_{1bar}}{R_P})\frac{k_B}{\Delta E_{ac}}
\end{equation}
We identify the two different transitions we use as thermometers by looking at the PL spectra of the NP in a cryostat at 6.5K (see Appendice Fig.\ref{fig:cryo_spectrum}). We verify the model in Eq.\ref{model_dI} and our choice of transitions by measuring the ratio of the Purple and Yellow transition at different temperatures in our confocal cryostat setup. The results are plotted in Fig.\ref{fig:spectrum_calib_cryo}b.\\
We note that in our samples the mean fluorescence wavelength seems to be a poor meter for the temperature in contrast with what has been previously observed \cite{luntz-martin_laser_2021}.

\begin{figure*}
	\centering
    \includegraphics[width=1\textwidth]{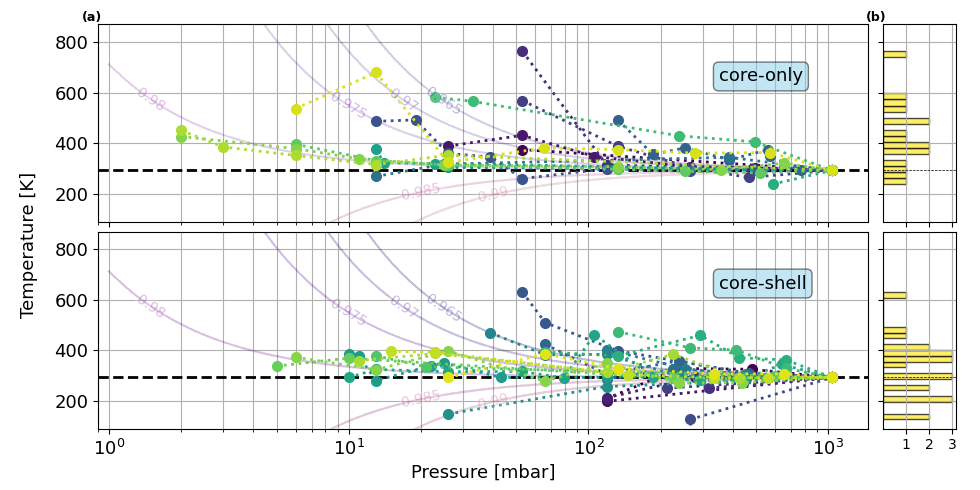}
    \caption{(a) Temperature dependence with pressure for different designs of nanocrystals. The different colours denote different particles. (top) core only 10$\%$Yb$^{3+}$:NaYF$_4$ and (bottom) core-shell 10$\%$Yb$^{3+}$:NaYF$_4$@NaYF$_4$. The solid lines represents the simulated internal temperature for different $\eta$ values(see Appendices). (b)We show an histogram (bin width 15K) of the temperature reached at the lowest pressure for each particle, we clearly see the superior thermodynamic properties of the core-shell (bottom) vs the bare nanoparticles (top).}
	\label{fig:T_vs_P_n_sim}
\end{figure*}

\section{Results}
\subsection{Thermodynamics of a levitated nanocryostat}
\begin{figure*}
	\centering
    \includegraphics[width=1\textwidth]{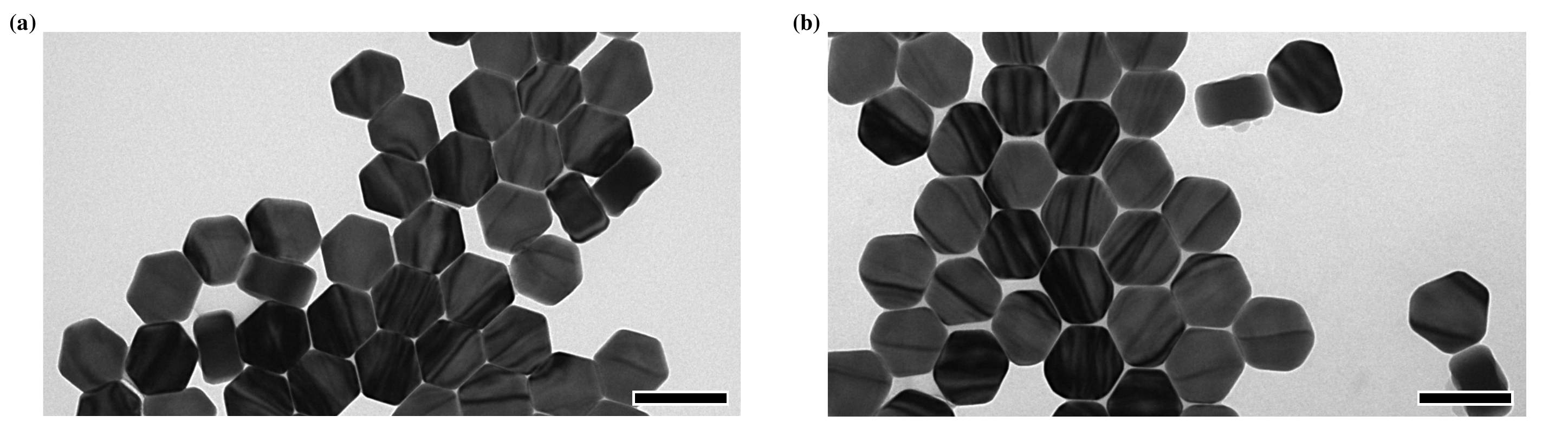}
    \caption{TEM of both nanocrystals. a) core and b) core-shell. Scale bar is 200 nm.}
	\label{fig:TEM}
\end{figure*}

In this study, we are interested in evaluating the cooling performance of two variants of 10$\%$ Yb$^{3+}$:NaYF$_4$ nanocrystals, with and without a 5-nm inert shell (see Fig. \ref{fig:TEM}). The intrinsic high quantum yield $\eta_e \approx 0.99$ of Yb$^{3+}$ ions makes them a candidate of choice for cooling purposes, as most of the absorbed laser radiation is re-emitted through anti-Stokes fluorescence. Non-radiative energy losses and parasitic absorption in the host crystal can however reduce or even quench the cooling efficiency. In upconversion nanocrystals (and in nanoparticles in general), non-radiative energy losses are mainly mediated through inter-ion energy transfer and finally loss at the surface. By growing an inert shell around the nanocrystal \cite{ren_quantifying_2021,mulder_understanding_2023} one can quench these processes thereby maximising the external quantum yield and thus the cooling power of the nanocryostat.\\
We monitor the internal temperature of the levitated nanocrystals when decreasing the pressure in the vacuum chamber. At ambient pressure, the nanoparticle thermalizes mostly through gas collisions while under moderate vacuum conditions (P $\leq$ 100 mbar) competition between laser absorption and fluorescence will define the final internal temperature.\\
The net cooling power in this regime can be expressed as $P_{cool}=\dot{Q}^{laser}_{heat}+ \dot{Q}^{fluo}_{cool}$ where heating through absorption of the pump laser (and trapping laser if they are not the same) is given by:
\begin{equation}
    \dot{Q}^{laser}_{heat} = (\alpha + \alpha_b)I
\end{equation}
with $\alpha$ denoting the absorption coefficient for the Yb$^{3+}$ ions and $\alpha_b$ the background absorption coefficient of the host nanocrystal. Using a four-level model \cite{jin_adaptive_2021} with a temperature-independant absorption, cooling through anti-Stokes fluorescence can be described by:
\begin{equation}
    \dot{Q}^{fluo}_{cool} = -\eta_e \frac{\omega_{f}}{\omega_{p}} \alpha I
\end{equation}
where $\eta_e$ is the external quantum yield, $\omega_f$ the mean fluorescence angular frequency and $\omega_p$ the pump laser. For a more detailed modelling, one would need to take into account the temperature dependence of both $\alpha$ and $\omega_f$.\\
Finally considering the thermalization with surrounding gas (in the Knudsen regime, $Kn \geq 10$), we can write:
\begin{equation}
    \dot{Q}^{gas}_{heat/cool} = - a_{acc}\sqrt{\frac{2}{3\pi}}\pi r^2 v_{th} \frac{\gamma_{sh}-1}{\gamma_{sh}+1}(\frac{T_{int}}{T_{gas}}-1)P_{gas}
\end{equation}
where $a_{acc}$ is the thermal accommodation coefficient which denotes the fraction of gas molecules that thermalize with the nanoparticle temperature. $r$ is the radius of the particle, $v_{th}$ is the mean thermal velocity of impinging gas molecules and $\gamma_{sh}=7/5$ is the specific heat ratio of a diatomic gas. For pressure above 100 mbar ($Kn \leq 10$ with $Kn \approx 1$ at ambient pressure), the mean free path $\lambda_{mfp}$ of gas molecules shortens and we can model the thermal exchange rate between the particle and the surrounding gas using \cite{liu_heat_2006}:
\begin{equation}
    \dot{Q}^{gas}_{heat/cool} = -\frac{8\pi r^2 k_g}{2r+\lambda_{mfp}G}(T_{int}-T_{gas})
\end{equation}
The factor G is given by $(18\gamma_{sh}-10)/a_{acc}(\gamma_{sh}+1)$. We note that both models (Knudsen regime and intermediary) give very close results for the range of parameters explored here.\\
In the present work, we are in the regime ($P_{gas} > 1$ mbar) where blackbody radiation contribution is negligible, so the heat equation for the system is simply given by:
\begin{equation}
    mc_v\frac{dT_{int}}{dt}=\dot{Q}^{gas}_{hc}+\dot{Q}^{laser}_h+\dot{Q}^{fluo}_c
\end{equation}
We can estimate the steady-state internal temperature of the levitated nanocrystal at different pressures assuming thermal equilibrium. By measuring the temperature for varying pressure and/or laser power we can then estimate the cooling efficiency of the nanocrystals.
\\
The results are compiled in Fig.\ref{fig:T_vs_P_n_sim}. One can assess the heating/cooling rate by looking at the change in temperature when varying the optical power for pressure below 500 mbar, as the thermal exchange with the gas reduces. The thermodynamic performance of the nanocrystals becomes apparent while reducing the pressure, when competition between absorption and fluorescence becomes the dominant thermalization mechanism. From the data, it appears that we cannot bring all the nanocrystals to lower pressures. It is expected that at the range of pressures explored here, a change in the NP temperature can induce strong photothermal forces that often lead to the ejection of the particle from the trap.
We found that 6 out of 22 core-shell (CS) NPs show cooling for only 2 (out of 16) for the bare nanoparticles. We measure a minimum temperature of 126 $\pm$ 2 K for a core-shell design at a pressure of 266 mbar. We also measure a CS nanoparticle cooling down to 148 $\pm$ 4 K down to a pressure of 26 mbar, hence for the first time bringing down the internal temperature of a levitated nanoparticle in the underdamp regime, opening up the possibility for the absolute cooling (internal and COM temperature) of a levitated nanoparticle.\\
Despite the nanoparticles being uniform in size/shape and PL, they show a variety of thermodynamic properties that can be reproduced in simulation by slight variations in quantum yield $\eta_e$ and/or background absorption $\alpha_b$. These two quantities are related as a stronger background absorption will mean a lower quantum yield.
Statistically, the core-shell nanoparticles have higher quantum yield, hence cooling power. It is worth noting that a mere change of 0.05$\%$ can mean going from a cryostat to a heater. From this study, we can deduce that solid-state refrigeration seems to require more stringent requirements than upconversion imaging. Indeed both sample of nanoparticles shows similar properties in shape, size and brightness but their cooling properties are clearly not as uniform. The fact that the core-only NP show barely any cooling seem to indicate that their properties were degraded. Furthermore, because we still measure cooling with the core-shell NPs, it seem reasonable to points towards a degradation of the surface of the NPs maybe during the surface modification step. As one can expect a bigger nanocrystal will yield more cooling (or heating) power for a fixed $\eta_e, \alpha_b$ and so it could be advantageous in the future to work with bigger mesoscopic particle.\\
The single particle thermodynamics presented here also show an interesting new avenue to characterise the quantum yield of levitated nanocryostat. 

\section{Conclusions}
In conclusion, we have characterized the thermodynamic (i.e. laser refrigeration) properties of a range of 10$\%$ Yb$^{3+}$:NaYF$_4$ nanocrystals, with and without an inert shell. Although nanoengineering brighter NP is relatively mature in the field of upconversion imaging, we here show for the first time that the same techniques can be employed to design more efficient nanocryostats. It opens avenues for more improvement: higher doping rate, co-doping with other lanthanide species, alternative designs such as core-multi-shell and even different aspect ratios. Achieving this goal will greatly benefit the levitodynamics community by offering a unique method to control the internal temperature of levitated objects, which will affect their oscillator coherence properties in UHV. Although the context of this work has been done with application in levitated optomechanics, the results and methods can be extended for instance to the field of biology where controlling the temperature of physiological media can add important capability.
\\
\section*{Acknowledgements}
C. L. is supported by the Sydney Quantum Academy Postdoctoral Research Fellowship and would like to thank Gabriel Hetet, John G. Bartholomew, Philippe Goldner, George Winstone and generally the LEVINET collaboration network for inspiring and insightful discussions. T. V. and R. P. R. acknowledge support from the Australian Research Council Centre of Excellence for Engineered Quantum Systems (Grant No. CE170100009) and Lockheed Martin.
The authors would like to acknowledge help in the preliminary stage of this experiment from Katherine Kinder and Michael Robinson as well as Xianlin Zheng for preparing the first generation nanocrystals samples.

\bibliography{REI_lev_nano_refrigeration.bib}

\section*{Appendices}
\subsection{Thermodynamics}
Here are the value used for the simulated thermodynamic performance (solid lines in Fig.\ref{fig:T_vs_P_n_sim}): $\alpha_b = 0.001\alpha$ with $N_t = 1.47\times10^{22}$ cm$^3$ and $\sigma = 1.80\times10^{-20}$ cm$^{-2}$ \cite{wiesholler_ybnder-doped_2019}, $a_{acc}$ = 0.05, $\overline{\lambda}_{fluo}$ = 999.6 nm.
\\
When the pressure is low enough (usually for $P_{gas} < 10^{-6}$ mbar), the blackbody radiation of both the environment and the particle starts to play a role and its contribution is given by \cite{chang_cavity_2010} :
\begin{equation}
    \dot{Q}^{bb}_h = \frac{24\xi_R(5)}{\pi^2\epsilon_0c^3\hbar^4}\alpha^{''}_{bb}k_B^5(T_{env}^5-T_{int}^5)
\end{equation}

\subsection{Thermometry}
\begin{figure}[H]
    \includegraphics[width=0.5\textwidth]{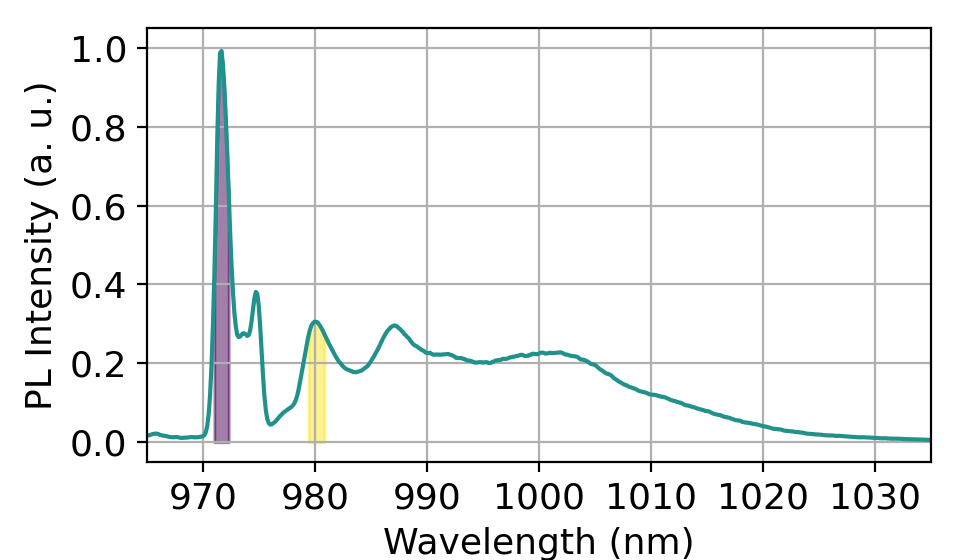}
    \caption{PL spectra of a core nanoparticle at 6.5K. The purple and yellow regions are the transition used for thermometry.}
	\label{fig:cryo_spectrum}
\end{figure}
We can potentially estimate the temperature by looking at the change in the linewidth of the main optical transition at 973 nm. In the range of temperature from 50 to 300 K, a nearly quadratic ($\approx T^{1.9\pm0.1}$) dependence of the homogeneous linewidth has been commonly reported in a large variety of host and lanthanide ions \cite{brundage_low-temperature_1986,lei_determination_1998}. For this range of temperatures, the broadening is dominated by the homogeneous optical linewidth of the ions. The phonon processes are the dominant causes of the broadening.

\subsection{Materials synthesis}
We here describe the synthesis for the two different types of nanocrystals discussed in the article: 10$\%$Yb$^{3+}$:NaYF$_4$ nanocrystals (diameter of 160 nm and thickness of 80 nm and 10$\%$Yb$^{3+}$:NaYF$_4$@NaYF$_4$ with a 5 nm inert outer shell (170 nm and thickness of 90 nm).\\
\textit{Chemical and reagents}: Ytterbium (III) chloride hexahydrate (YbCl$_3$$\cdot$6H$_2$O, 99.99$\%$),  yttrium (III) chloride hexahydrate (YCl$_3$$\cdot$6H$_2$O, 99.99$\%$), oleic acid (OA, 90$\%$), 1-octadecene (ODE, 90$\%$), sodium oleate ($\geq$82, fatty acids), ammonium fluoride (NH$_4$F, $\geq$98$\%$), and sodium hydroxide (NaOH, $\geq$97$\%$, pellets) were purchased from Sigma-Aldrich and used as received without further purification.\\
\textit{Synthesis of core nanoparticles}: The growth of core nanoparticles was precisely controlled by a purpose-built automated growth system \cite{ren_quantifying_2021} To synthesize $\beta$-NaY$_{0.9}$Yb$_{0.1}$F$_4$ core particles, YCl$_3$ and YbCl$_3$(total 1 mmol) powder were dissolved in methanol and mixed in a three-neck flask with 6 mL OA and 15 mL ODE. The mixture was stirred and heated to 75 °C for 30 min and then 160 °C for another 30 min. After cooling back to room temperature, 2.5 mmol NaOH and 3 mmol NH$_4$F were dissolved in methanol and added to the flask. The solution was kept at 75 °C and 160 °C for 30 min, respectively. Then the temperature was increased to 310 °C for 90 min. All the reaction was carried out under Argon gas flow. After the solution was cooled down to room temperature, the nanoparticles were washed and precipitated with ethanol, collected by centrifugation (6000 rpm for 6 min), and dispersed in cyclohexane.\\
\textit{Synthesis of the inert shell precursor}: The synthesis procedure of shell precursor (NaYF$_4$) was similar to the core nanoparticles except using YCl$_3$ instead of YbCl$_3$. Moreover, the reaction was stopped before the solution was heated to 310 °C.\\
\textit{Synthesis of the core-shell nanoparticles}: Based on the diameter of the core and the intended shell thickness, 1.5 mL NaYF$_4$ precursor and 70 mg core nanoparticles were mixed in a three-neck flask with 6 mL OA and 15 mL ODE. The mixture was stirred and heated to 75 °C for 30 min and then to 310 °C for 40 min. The process was under the protection of Argon gas flow. After the solution was cooled down to room temperature, core-shell nanoparticles were washed and dispersed by the same method as the core particles.\\
\textit{Modification of the surface for dilution}: We need to dilute the nanoparticle solution in ethanol before nebulization into the trap so we modify the surface from hydrophobic to hydrophilic. 2 mg UCNPs dissolved in 1 mL cyclohexane were mixed with 9 mL pH 4 hydrochloride acid solution. The mixture was vortexed for 2 h. The hydrophilic nanoparticles were collected by centrifugation (8000 rpm for 15 min) and dispersed in milli-Q water.\\

% \nocite{*}

% Produces the bibliography via BibTeX.

\end{document}